\documentclass[final,5p,times,twocolumn,authoryear]{elsarticle}

\usepackage{amssymb}
\usepackage{amsthm}
\usepackage{amsmath}
\usepackage{booktabs}       % professional-quality tables
\usepackage{caption}
\usepackage{hyperref}       % hyperlinks
\hypersetup{
    colorlinks=true,
    linkcolor=cyan
    }

\usepackage{tikz}
\usetikzlibrary{positioning}
\usepackage{tikzscale}
\usepackage{makecell}

\usepackage{lineno}
\journal{Marine Mammal Science}

\begin{document}

\begin{frontmatter}

\title{Advancing Marine Bioacoustics with Deep Generative Models: A Hybrid Augmentation Strategy for Southern Resident Killer Whale Detection}

\author[inst1]{Bruno Padovese\corref{cor1}}
\ead{bruno_padovese@sfu.ca}
\cortext[cor1]{Corresponding author}

\affiliation[inst1]{organization={School of Environmental Science},%Department and Organization
            addressline={Simon Fraser University}, 
            city={Burnaby},
            postcode={V5A 1S6}, 
            state={BC},
            country={Canada}}

\affiliation[inst2]{organization={Faculty of Computer Science},%Department and Organization
            addressline={Dalhousie University}, 
            city={Halifax},
            postcode={B3H 1W5}, 
            state={NS},
            country={Canada}}

\affiliation[inst3]{organization={Department of Mathematics and Statistics},%Department and Organization
            addressline={Dalhousie University}, 
            city={Halifax},
            postcode={B3H 1W5}, 
            state={NS},
            country={Canada}}

\author[inst1,inst2]{Fabio Frazao}
\author[inst3]{Michael Dowd}
\author[inst1]{Ruth Joy}

\begin{abstract}

Automated detection and classification of marine mammals vocalizations is critical for conservation and management efforts but is hindered by limited annotated datasets and the acoustic complexity of real-world marine environments. Data augmentation has proven to be an effective strategy to address this limitation by increasing dataset diversity and improving model generalization without requiring additional field data. However, most augmentation techniques used to date rely on effective but relatively simple transformations, leaving open the question of whether deep generative models can provide additional benefits. In this study, we evaluate the potential of deep generative for data augmentation in marine mammal call detection including: Variational Autoencoders, Generative Adversarial Networks, and Denoising Diffusion Probabilistic Models. Using Southern Resident Killer Whale (\textit{Orcinus orca}) vocalizations from two long-term hydrophone deployments in the Salish Sea, we compare these approaches against traditional augmentation methods such as time-shifting and vocalization masking. While all generative approaches improved classification performance relative to the baseline, diffusion-based augmentation yielded the highest recall (0.87) and overall F1-score (0.75). A hybrid strategy combining generative-based synthesis with traditional methods achieved the best overall performance with an F1-score of 0.81. We hope this study encourages further exploration of deep generative models as complementary augmentation strategies to advance acoustic monitoring of threatened marine mammal populations. 

\end{abstract}


\end{frontmatter}

% \linenumbers

\section{Introduction}

Killer Whales (\textit{Orcinus orca}) are a widespread species inhabiting diverse marine environments across all oceans. The species comprises several distinct ecotypes, some of which can be further divided into populations with each adapted to specific ecological niches and characterized by unique behaviors and intricate social structures. Their extensive vocal repertoire spans frequencies from 0.5 to 100 kHz and includes three primary call types: echolocation clicks, single-toned whistles, and discrete pulsed calls \citep{ford1987catalogue, ford1978underwater}, each serving distinct behavioral functions. Echolocation clicks, concentrated between 20 and 100 kHz, are primarily used for navigation and prey detection \citep{barrett1996mixed, au2004echolocation}. Whistles (0.5–25 kHz) are primarily used for social interactions within pods \citep{miller2006diversity, thomsen2001characteristics, riesch2008whistle}. Pulsed calls, which occur in the same frequency range, are the most common vocalization type, supporting group cohesion, individual identification, and more complex social communication \citep{ford1987catalogue, ford1978underwater, filatova2009usage}. Furthermore, population-specific vocal dialects add a layer of complexity to their acoustic repertoire, reflecting the social and cultural divergence that distinguishes populations within and across ecotypes.

In the Northeast Pacific, Killer Whales have evolved into three genetically and culturally distinct ecotypes \citep{barrett1996mixed, morin2024revised}, which frequently share overlapping habitats: Resident, Transient, and Offshore Killer Whales \citep{ford1998dietary, riesch2012cultural, baird1988variation}. The Resident (fish-eating) ecotype contains two populations, the Northern Resident population that is made up of three linguistic clans, and the Southern Resident population which consists of one linguistic clan. The populations remain reproductively isolated \citep{morin2024revised, riesch2012cultural}, and there are strong cultural and linguistic barriers between clans. The Southern Resident Killer Whale population (SRKW) also known as J-clan, is made up of three stable, matrilineal family groups, identified as J, K and L pod. The SRKW range stretches from California to southeast Alaska, and with heavy use in the cross-boundary waters of the Salish Sea. The population is listed as endangered in Canada and is protected by the Species at Risk Act (SARA) \footnote{\href{https://species-registry.canada.ca/index-en.html\#/species/699-5}{https://species-registry.canada.ca/index-en.html\#/species/699-5}} and by the Endangered Species Act \footnote{\href{https://www.fisheries.noaa.gov/topic/laws-policies/endangered-species-act}{https://www.fisheries.noaa.gov/topic/laws-policies/endangered-species-act}} in the US. By the end of 2024, The Center for Whale Research, the organization responsible for semiannual SRKW census numbers, estimated there were only 75 individuals remaining \footnote{\href{https://www.whaleresearch.com/whalereport25}{The Center for Whale Research Report}}.

In light of this critical population status, there is significant interest from researchers, citizen scientists, government agencies, and First Nation groups in the protection and conservation of SRKW and their habitat. In many cases, visual sightings remain the primary method of monitoring these individuals \citep{olson2018sightings, sato2021southern}, but this approach is constrained by weather conditions and the need for active, daytime surface observations, which are predominantly conducted by citizen scientists. More recently, the availability of affordable acoustic recording devices and expanded data storage capacity has enabled large-scale passive bioacoustic monitoring, leading to the creation of extensive audio datasets \citep{roch2017organizing, dede2014long, webster2017sound}. However, a lack of highly specialized person-hours for bioacoustic data analysis creates a significant bottleneck in the manual detection, annotation, and validation of whale calls across large acoustic datasets. This is a time-consuming process that can span weeks or months for a single hydrophone deployment. Consequently, there is a need for efficient processing methods that can automate key steps in the workflow, such as vocalization classification and detection, within a reasonable timeframe \citep{stowell2022computational} of a few hours or days instead of weeks. This demand for automated solutions has driven the development of SRKW detection and classification algorithms based on signal processing techniques \citep{gillespie2013automatic, shapiro2009versatile} and machine learning (ML) \citep{sharpe2019call, brown2010automatic}. Within ML, deep learning (DL) approaches, originally developed for applications in image, speech, and music processing \citep{abesser2020review, opensourceseparation:book} have gained traction due to their success in complex pattern recognition and feature extraction \citep{lecun2015deep, Goodfellow-et-al-2016}. Deep Neural Networks (DNN) have been shown to outperform traditional ML techniques \citep{stowell2022computational, morfi2021deep}, including in the context of Killer Whale acoustic detection and classification \citep{bergler2019orca, hauer2023orca, bergler2021fin}.

Two-dimensional spectrograms, typically segmented into fixed-length audio clips (e.g., 1-second or 10-second duration), are commonly used as inputs to DNNs for cetacean classification and detection tasks \citep{kirsebom2020performance, bergler2019orca, li2020learning}. This practice is also commonplace across other fields such as sound event classification \citep{ozer2018noise}, bird song recognition \citep{kahl2021birdnet}, as well as speech \citep{wang2018supervised} and music classification \citep{elbir2020music}. In marine bioacoustics, spectrograms are also one of the primary tools for visualizing and analyzing acoustic data, allowing researchers to quickly identify patterns that may be missed through manual listening, especially for sounds outside the human audible range. Furthermore, many acoustic signals, such as Killer Whale vocalizations, contain frequency-modulated (FM) components that are discernible in spectrograms \citep{rabiner1993fundamentals}, making them suitable for use in automated classification models.

However, while spectrogram-based DNNs have shown promise \citep{kirsebom2020performance, bergler2019orca, shiu2020deep}, their effectiveness is constrained by the availability of sufficient high-quality annotated data \citep{priestley2023survey, gudivada2017data}, a common issue in marine bioacoustics. As mentioned above, obtaining high-quality annotations is challenging and expensive, particularly for marine environments where target vocalizations can be sparsely distributed against a backdrop of overwhelming underwater environmental noise \citep{bergler2019orca, stowell2022computational, padovese2023adapting}. This scarcity of annotated data represents a critical bottleneck in the development of effective DL models for marine mammal classification. Therefore, before investing in costly and time-consuming efforts to manually annotate and assign labels to acoustic data, an effective strategy to mitigate this limitation is to use data augmentation to artificially enhance the diversity of small bioacoustic training dataset \citep{stowell2022computational, li2021automated}.

Data augmentation has long been used to address challenges associated with small or unbalanced datasets by creating modified versions of existing data \citep{shorten2019survey}. These augmented samples are variations of the original recordings that were not present in the training set but are theoretically possible within the same context. Generally, augmentation methods can be divided into two categories: ``na\"ive" augmentations and data-based augmentations. Na\"ive augmentations are characterized by their simplicity and do not take the specific problem being addressed into consideration. In bioacoustics, commonly used na\"ive methods include time-shifting \citep{shiu2020deep}, time and frequency masking \citep{park2019specaugment}, and noise addition \citep{mishachandar2021diverse}. These methods are popular due to their ease of implementation, effectiveness in improving model performance, and ``safety”, meaning they are unlikely to compromise the meaning of the audio, ensuring the augmented signals remain consistent with the original class \citep{stowell2022computational}.

In contrast, data-based augmentations leverage domain knowledge to transform samples based on the characteristics of the environment and target species. These can include simple operations like sound mixing \citep{padovese2021data} or more complex techniques like sound propagation modeling, which simulates how sound is distorted as it propagates through water \citep{binder2018impacts}. Some studies have experimented with more risky transformations, including warping \citep{park2019specaugment}, pitch shifting \citep{li2021automated}, and time stretching \citep{li2021automated}. While these methods can potentially increase the diversity of the training data, they also risk distorting subtle acoustic features that are unique to specific species or call types \citep{stowell2022computational}. As a result, the appropriate choice of augmentation techniques is highly context-dependent and should be tailored to the characteristics of each dataset and species.

While traditional data augmentation helps address the limitations of small or unbalanced datasets, it is inherently constrained by the nature of the transformations themselves, which can only recombine or distort existing information in limited ways. To further expand the dataset and introduce new variability, data synthesis that relies on algorithms to generate artificial data has emerged as a valuable tool \citep{king2014role, reichert2015noise}. Already widely used in marine mammal communication research \citep{king2014role, reichert2015noise}, synthetic data can also serve as artificial training examples to supplement or even replace real-world data in machine learning models \citep{li2020learning}. Importantly, synthetic data generation and classical data augmentation strategies are not mutually exclusive; combining both can yield superior results in deep learning applications by maximizing the diversity of training data.

Within data synthesis, DL-based generative models learn the distribution of the dataset's feature space to create new, unseen samples drawn from this distribution, introducing entirely new patterns not found in the original dataset. Generative models have long been used for tasks such as speech and music generation \citep{shorten2019survey}. Some studies have begun applying generative models in the field of bioacoustics; this field is still emerging, with limited research having made its way into the peer-review literature. Most research has focused on the generation aspect, typically using Generative Adversarial Networks (GANs) \citep{goodfellow2014generative} or Variational Autoencoders (VAEs) \citep{kingma2013auto} to operate on time-frequency spectrograms \citep{zhang2022dolphin, nieto2024soundscape}. Although a handful of studies have also applied generative models specifically for data augmentation \citep{herbst2024empirical, li2023learning}, even fewer have systematically compared these methods to traditional data augmentation approaches.

Recently, denoising diffusion probabilistic models (DDPMs) \citep{ho2020denoising} have gained attention for their ability to generate higher-quality samples compared to their GAN or VAE counterparts \citep{NEURIPS2021_49ad23d1}. Unlike these earlier approaches, DDPMs generate high-quality, diverse samples by gradually refining noise into informative data, offering better stability compared to traditional GANs or VAEs. These models have shown significant potential for producing realistic synthetic data and have been studied for improving neural network training in tasks such as image classification \citep{trabucco2023effective}. In bioacoustics, diffusion models remain largely unexplored, with, to the best of our knowledge, few studies investigating their potential for this purpose \citep{herbst2024empirical}.

In this work, an aim is to extend the application of DDPMs for data augmentation tailored to SRKW vocalizations. Our approach assumes that, similar to traditional augmentation, additional latent information can be derived from the original dataset through the use of deep generative models. By combining generative models with simpler augmentation techniques, we aim to improve the overall performance of deep learning classifiers, while overcoming the limitations of each method when used independently, particularly under conditions of scarce annotated data.

Our work makes three key contributions: (i) we introduce the first application of diffusion models for data augmentation of SRKW vocalizations; (ii) we propose a hybrid approach integrating diffusion-based synthetic data generation with traditional augmentation techniques to enhance dataset diversity and model robustness; and (iii) we conduct a comparative evaluation of generative versus traditional data augmentation methods, addressing a gap in the literature by systematically assessing their impact on classification performance.

\section{Materials and Methods}
\label{sec:methods}

\subsection{Datasets}
\label{sec:data}

\begin{figure}[ht]
    \centering
    \includegraphics[trim=3pt 0pt 0pt 0pt, clip, width=0.9\linewidth]{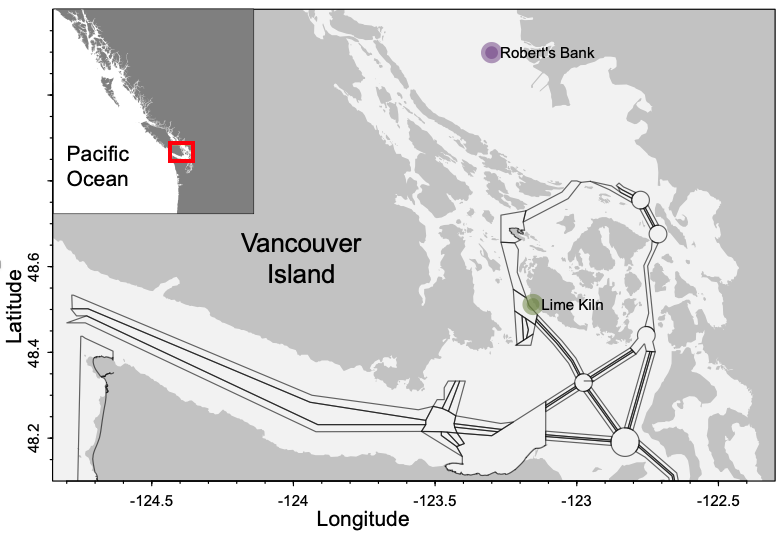}
    \caption{Locations of the two hydrophone deployments in the Salish Sea (Lime Kiln and  Roberts Bank). The commercial shipping lanes are shown as black lines.}
    \label{SalC}
\end{figure}

The underwater acoustic data used in this study were collected from two hydrophone deployments in the Salish Sea, off the west coast of North America (Figure \ref{SalC}). We focused on discrete pulsed calls from SRKW pods J, K, and L, with all vocalizations manually verified as true positives. Non-tonal sounds, such as whistles and echolocation clicks, were excluded. The recordings were obtained using different systems and protocols, each with distinct site characteristics, as detailed below.

The first dataset was collected near Lime Kiln State Park, off San Juan Island ($48^\circ 30' 42''\ \mathrm{N},\ 123^\circ 09' 15''\ \mathrm{W}$), through deployments conducted by SMRU Consulting Ltd. This dataset includes 1,633 audio files, each lasting one minute, recorded between August 29, 2018, and October 16, 2019. The recordings were captured using a hydrophone deployed at a depth of approximately 23 meters and sampled at 250 kHz. To identify Killer Whale vocalizations, recordings were first processed using the PAMGuard whistle and moan detector \citep{gillespie2008pamguard}, which generated initial binary detections indicating the presence or absence of potential biological sounds. All detected files were then manually reviewed in full, and vocalizations from various marine mammal species and ecotypes were annotated. From these, only annotations corresponding to confirmed KW calls were retained for this study, resulting in a total of 1,261 annotations from 200 files. This dataset was used exclusively for model training and hyperparameter tuning.

The second dataset was collected at Robert's Bank, British Columbia ($49^\circ 01' 07.8''\ \mathrm{N},\ 123^\circ 11' 32.8''\ \mathrm{W}$), by JASCO Applied Sciences. It comprises 1,562 5~minute audio files, recorded between September 21, 2015, and April 12, 2018. Recordings were made using a hydrophone deployed at a depth of approximately 168 meters and sampled at 64 kHz. Killer Whale encounters were initially identified using a proprietary detection algorithm developed by JASCO Applied Sciences. These encounters were then manually reviewed and annotated by expert analysts for the presence of Killer Whale vocalizations. For this study, we selected only those annotations confirmed to originate from KWs, resulting in 1,263 annotations across 22 files. These files and annotations were used exclusively for testing.

% The third dataset was collected at East Point by the Saturna Island Marine Research and Education Society (SIMRES). It consists of 143 audio files, each five minutes in duration, recorded between June 24, 2022, and October 3, 2022. Recordings were made near the commercial shipping channel in Boundary Pass, using a hydrophone deployed at a depth of approximately 18 meters and sampled at 128 kHz. Unlike the previous datasets, these recordings were not pre-processed with any automated detection algorithms. Instead, all files were manually reviewed and annotated for killer whale vocalizations. Each annotated signal was assigned a confidence rating of ‘low’, ‘medium’, or ‘high’, reflecting the annotator’s certainty. Calls that could not be confidently identified were labeled as ‘unknown’. For this study, we retained only annotations classified as SRKW vocalizations with a confidence rating of ‘medium’ or ‘high’, resulting in 2,025 annotations across 102 files.

The annotations were made publicly available through the HALLO (Humans and Algorithms Listening and Looking for Orcas) project. The full set of annotations can be accessed via the project’s GitHub repository.\footnote{\url{https://github.com/coastal-science/hallo-data/tree/main}} The underlying audio data were originally released in \cite{palmer2025public}.

\subsubsection{Data Preparation}
\label{sec:data-prep}

Killer Whale audio segments were extracted from recordings according to the annotations. Each labeled segment was isolated as a 3-second audio clip, a duration long enough to capture most SRKW calls \citep{FRAZAO2025103297}, while short enough to avoid overwhelming the neural network with excessive background noise. For the background class, which contains only (non-KW) background noise, random segments were drawn from the recordings while avoiding overlap with annotated regions. All recordings were downsampled to 24 kHz with anti-aliasing to ensure consistent processing across datasets with original sampling rates ranging from 64-250 kHz. This cutoff frequency was chosen as the resulting Nyquist frequency (12 kHz) fully encompasses the fundamental frequencies and harmonics characteristic of SRKW calls \citep{ford1989acoustic} while avoiding unnecessary computational overhead from higher sampling rates. To create a balanced dataset, the number of background segments randomly selected was made equal to the number of SRKW clips, and both were combined to form the complete baseline training dataset.

\subsubsection{Spectrogram Computation}
\label{sec:speccomp}

We computed 128-band Mel spectrograms from the 3-second segments derived from the annotation windows (Section~\ref{sec:data-prep}). Spectrograms were generated using a 50 ms Hann window (NFFT of 1200 samples) with a 12.5 ms hop length (300 samples), representing a 75\% overlap at the 24 kHz sampling rate. This configuration produced Mel spectrograms with frequency coverage up to 12 kHz, adopting a similar parameterization that has been successfully employed in SRKW classification tasks \citep{FRAZAO2025103297}. Amplitude values were then converted to a decibel scale for subsequent neural network processing.

\subsection{Time-Shifting Augmentation}
\label{sec:time-shift}

Time-shifting artificially expands the dataset by temporally displacing audio clips within their original recordings. For a given spectrogram \( x \) of duration $T$, we generate $N$ augmented samples by shifting the annotation window forward and backward in increments of $0.5$ seconds. Each shifted window maintains a minimum overlap of $50$\% with the original annotation, ensuring that the shifted segments remain contextually relevant to the labeled content without introducing mismatched examples. This guarantees that every annotation is represented by at least $N\geq1$ instances, though the exact value of $N$ depends on the original annotation duration and the overlap requirement. Beyond simply increasing sample size, time-shifting improves model robustness to natural temporal variations in audio events. Unlike methods such as noise addition \citep{white2022more} or random masking \citep{park2019specaugment}, it preserves the original acoustic features while maximizing the value of the limited labeled data.

\subsection{Vocalization Mask Augmentation}
\label{sec:masking}

While most augmentation strategies target within-class variability, such as differences in pitch, duration, or timing, they often overlook the variability introduced by natural soundscapes. This contextual information, such as various sources of transient sounds and shifting ambient noise conditions, can influence how a DNN distinguishes vocalizations from background events. Ignoring this context can limit model generalization, particularly in real-world deployments where acoustic conditions vary widely. To address this and create realistic, context-aware synthetic vocalizations, we developed a vocalization mask augmentation strategy based on high-quality SRKW call examples. 

To create the masks, we used a curated catalogue of high-quality SRKW vocalizations as the source for clean signal references. The samples in this collection were compiled over several decades of research by Dr. John Ford and made publicly available\footnote{\url{https://orca.research.sfu.ca/call-library/home.html?v=20240530-1727}}. We first computed spectrograms from the recordings following the procedure described in Section~\ref{sec:data-prep}. We then projected the spectrograms into a lower-dimensional space using PCA. In this representation, the first principal component tends to capture the broad scale pattern that is consistent across time and frequency, and explain the most variance in the dataset. These typically correspond to persistent background noise and non-whale acoustic features present in the environment. By subtracting this component from each original spectrogram, we can obtain a mostly denoised representation that emphasized the vocalization while suppressing background content.

\begin{figure*}[ht]
    \centering
    \includegraphics[width=0.8\textwidth]{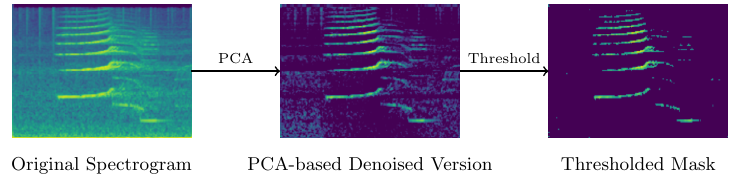}
    \caption{Overview of the vocalization mask construction process. Starting from the original spectrogram (left), we first apply PCA-based background subtraction to emphasize vocalization components (center). A subsequent percentile-based thresholding step produces a sparse, high-contrast mask (right) that preserves the primary vocal features while suppressing residual background noise.}
    \label{fig:mask_pipeline}
\end{figure*}

Next, to further refine these masks, we applied a thresholding step in which all pixel values below the $i$-th percentile of the spectrogram's dynamic range were set to zero. This process suppressed residual background noise resulting in sparse, high-contrast masks that captured most vocal features. The full process of mask construction and refinement is illustrated in Figure~\ref{fig:mask_pipeline}.

The resulting masks were then linearly combined with randomly sampled background spectrograms from the Robert’s Bank dataset. Specifically, we selected segments from recordings that were outside the 22 files reserved for the test set and manually verified to contain only background noise. By combining the masks with actual environmental noise, we created new, high-fidelity vocalizations that reflect the acoustic complexity encountered in the field. This augmentation approach exposes the model to more realistic combinations of signal and background, ultimately improving its robustness and generalization to unseen acoustic environments.

In practice, however, the manual validation of background segments described above may not even be necessary. In real-world applications, randomly sampling unlabeled audio from long-term passive acoustic recordings is likely to overwhelmingly yield only background noise. This makes the approach both scalable and easy to implement.

\subsection{VAEs}
\label{sec:vae}

Variational Autoencoders (VAEs) \citep{kingma2013auto} are a class of generative models that learn to generate new samples by encoding inputs into compact representations and decoding them back into the original data format. This encoder–decoder structure allows the model to capture the essential features of the input data. The intermediate representation, commonly referred to as the latent space, serves as a compressed version of the data, which VAEs learn to organize in a smooth and continuous manner suitable for generation.

In a standard autoencoder, the encoder learns to compress an input \( x \) into a lower-dimensional representation \( z \) through successive layers, gradually compressing the information into a simplified form that captures its most important features. The decoder does the opposite, attempting to reconstruct the input as \( \hat{x} = p_\theta(x \mid z) \), where \( \hat{x} \) is the reconstructed input, and \( p_\theta(x \mid z) \) represents the decoder network. This network gradually expands the compact representation back into a full-resolution output. Crucially, the decoder can only learn to accurately reconstruct inputs if the encoded representation is informative and semantically meaningful. To achieve this, the model is trained to minimize the difference between the input and its reconstruction, encouraging it to preserve essential structure while discarding noise or redundancy. However, because the encoder only learns to handle inputs it has seen during training, the resulting latent space can be irregular and fragmented, even if it compresses data well. As a result, there's no guarantee that randomly sampling from this space will produce valid or meaningful outputs, making standard autoencoders poorly suited for generative tasks.

VAEs address this by introducing a probabilistic encoding scheme. Instead of mapping each input to a single fixed point in the latent space, the encoder \(q_\phi(z \mid x) \) learns to represent it as a Gaussian distribution defined by a mean vector \( \mu \) and a standard deviation vector \( \sigma \). A sample is drawn from this normal distribution using the so-called reparameterization trick:

\begin{equation}
z = \mu + \sigma \cdot \epsilon, \quad \epsilon \sim \mathcal{N}(0, I)
\end{equation}
where \( I \) denotes the identity covariance matrix, following the standard VAE formulation.

The decoder \( p_\theta(x \mid z) \) then reconstructs the input from this sampled vector. Unlike in standard autoencoders, where \( z \) might be sampled from  uninformative parts of the latent space, VAEs are explicitly trained to make their latent space well-organized and continuous. This is done by nudging the learned distributions \(q_\phi(z \mid x) \) to stay close to a known prior, such as a standard normal distribution \(\mathcal{N}(0, I)\). As a result, VAEs learns to fill the latent space smoothly, such that small changes in \( z \) correspond to gradual changes in the decoded output. This structured organization makes it possible to sample new, realistic data simply by drawing \( z \sim \mathcal{N}(0, I) \) and passing it through the decoder.

The model is trained to minimize the following loss \citep{kingma2013auto}:

\begin{equation}
\mathcal{L}_{\text{VAE}} = -\mathbb{E}_{z \sim q_\phi(z \mid x)} [\log p_\theta(x \mid z)] + \beta\ \text{KL}(q_\phi(z \mid x) \| p(z))
\end{equation}
where the first term is the reconstruction loss, encouraging the decoder to accurately reconstruct the input \( x \) from the sampled latent representation \( z \), and the second term is a Kullback–Leibler (KL) divergence that regularizes the encoder’s output distribution to remain close to the standard normal prior \( p(z) = \mathcal{N}(0, I) \). \( \beta \) is a weight that controls the strength of the regularization; in this work, it was set to 1.

The stability and ability of VAEs to capture a wide range of data distributions make them advantageous for data augmentation tasks \citep{shorten2019survey}. However, a common drawback of VAEs is that their outputs tend to be less sharp or detailed than the original inputs or compared to those produced by other generative models, such as GANs \citep{kingma2019introduction, wang2020survey}. This blurriness arises from the probabilistic nature of the decoder and the averaging effect of the reconstruction loss, which can smooth out fine-grained features. 

In this work, we employed a standard convolutional VAE architecture in which both the encoder and decoder are composed of stacked convolutional layers with batch normalization and ReLU activations. The model takes as its input 2D spectrograms. The generated spectrograms aim to match the overall distribution of the Lime Kiln training data. Further implementation details, including architectural specifications and training configuration, are provided in Section~\ref{sec:expSetup}.

\subsection{GANs}
\label{sec:gan}

Generative Adversarial Networks (GANs) are another category of generative models that learn to synthesize data by jointly training two competing neural networks: a generator \( G \), which produces synthetic samples intended to resemble those from the training distribution, and a discriminator \( D \), which attempts to distinguish between real and generated samples \citep{goodfellow2014generative}. The central idea behind GANs is that, as the discriminator improves at identifying fake samples, the generator must produce increasingly realistic outputs in order to fool it. Conversely, as the generator becomes better at producing convincing samples, the discriminator must also improve to maintain its ability to detect fakes. This dynamic coupling forms an adversarial feedback loop in which both networks iteratively enhance their capabilities. During training, the generator and discriminator are optimized simultaneously in a zero-sum game, where each network's objective directly opposes the other. Once the adversarial training process stabilizes, the generator can be used independently to synthesize new samples, whether images, or, in our case, spectrograms of SRKW vocalizations.
                
GANs have been widely applied in artificial image generation \citep{karras2019style, choi2018stargan}, particularly in domains such as facial image synthesis \citep{karaouglu2021self}, scene reconstruction \citep{wang2018high}, and image-to-image translation \citep{isola2017image}. While their application to time–frequency representations such as spectrograms is less common, some works have demonstrated the potential of GANs in the bioacoustic domain \citep{li2023learning, bergler2022orca, 9412721}. Unlike natural image synthesis, spectrogram generation of vocalizations involves fine, curvilinear features such as SRKW harmonic ridges, which are more sensitive to any distortion.

Formally, the generator network maps a random noise vector \( z \sim \mathcal{N}(0, I) \) to the data space, producing a synthetic spectrogram \(\hat{x} = G(z) \). The discriminator \( D(x) \) receives either a real spectrogram \( x \in X_{real} \), drawn from the data distribution \( p_{\text{data}} \), or a synthetic one \( \hat{x} \), and outputs a scalar representing the probability that the input is real. The networks are trained with opposing objectives defined by the minimax loss function:

\begin{equation}
\min_G \max_D \ \mathbb{E}_{x \sim p_{\text{data}}}[\log D(x)] + \mathbb{E}_{z \sim p(z)}[\log(1 - D(\hat{x}))].
\end{equation}

While GANs have demonstrated impressive capabilities in synthesizing visually realistic images, they present several limitations. First, GANs are notoriously difficult to train and highly sensitive to hyperparameter choices, network architecture, and optimization dynamics \citep{arjovsky2017towards}. These factors often lead to unstable training, non-convergent behavior, or poor-quality outputs \citep{agarwal2021detecting}. Second, GANs are prone to mode collapse \citep{che2016mode}, a failure in which the generator produces a limited set of outputs, such as repeatedly generating highly similar or identical samples \citep{srivastava2017veegan, mariani2018bagan, dhariwal2021diffusion}. This leads to a lower diversity of generated data compared to the real distribution, limiting the model's ability to represent the full variability of the training data, which is an important drawback when using GANs for data augmentation when training a classifier.

Following \citet{goodfellow2014generative}, a number of advancements have focused on improving training stability and sample quality. These include architectural refinements such as the Deep Convolutional GAN (DCGAN) \citep{radford2015unsupervised}, as well as alternative training objectives like the Wasserstein GAN (WGAN) \citep{arjovsky2017wasserstein} and its gradient-penalized variant, WGAN-GP \citep{gulrajani2017improved}. Ultimately, however, despite these improvements, GANs still suffer from fundamental training challenges and instability.

In this work, we adopted the DCGAN architecture, which introduces architectural constraints such as convolutional layers without fully connected components, batch normalization, and ReLU/LeakyReLU activations to improve training stability and sample quality. Our implementation follows the original GAN formulation described above, with both the generator and discriminator operating on 2D spectrogram tensors. The generator produces time–frequency representations that aim to match the distribution found in the Lime Kiln training set. The model was trained using the standard GAN loss. A detailed description of the network architecture and training routine is provided in Section~\ref{sec:expSetup}.

\subsection{DDPM}
\label{sec:DDPM}

Denoising Diffusion Probabilistic Models (DDPMs) are a class of generative models that synthesize data through an iterative denoising process \citep{ho2020denoising}. DDPMs operate in two phases: a forward process, which incrementally corrupts training samples (e.g., spectrograms) by adding Gaussian noise over discrete timesteps, and a reverse process, which learns to iteratively recover the original data by predicting and removing this noise \citep{nichol2021improved}. During training, the model learns the underlying structure of the data distribution by estimating the noise at each corruption step \citep{sohl2015deep}. During inference, the trained model ``hallucinates" novel samples (spectrograms in our case) by progressively denoising pure random noise. This reverse trajectory can generate realistic outputs that closely approximate the statistical distribution of the training dataset, enabling the creation of entirely new, yet plausible, synthetic vocalizations \citep{kong2020diffwave, herbst2024empirical}.

\begin{figure*}[ht]
    \centering
    \includegraphics[width=0.8\textwidth]{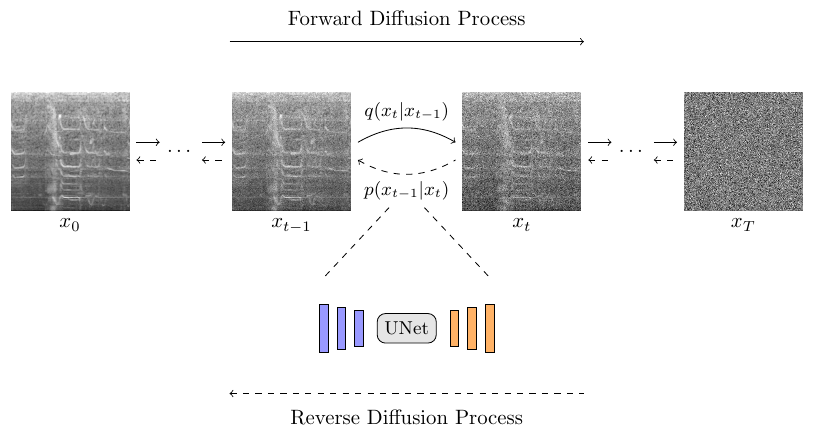}
    \caption{Illustration of a Denoising Diffusion Probabilistic Model (DDPM) for synthesizing spectrograms. The forward diffusion process (solid arrow) incrementally corrupts a clean input spectrogram $x_0$ into pure noise $x_T$ over $T$ timesteps. The reverse diffusion process (dashed arrow) learns to denoise by training a U-Net to predict and remove noise at each step.}
    \label{DDPM}
\end{figure*}

The forward process incrementally transforms a clean SRKW spectrogram $x_0=x$ into pure noise $x_T$ over $T$ timesteps (i.e., discrete diffusion steps) using a predefined noise schedule (Figure \ref{DDPM}). The noise schedule dictates how the noise is gradually added at each timestep through the noise schedule coefficients \( \alpha_t \). These coefficients \( \alpha_t \in (0, 1) \) determine the proportion of the original signal retained at each timestep, with smaller values of \( \alpha_t \) introducing more noise. The cumulative product of these coefficients, \( A_t = \prod_{s=1}^t \alpha_s \), determines the amount of the original clean signal \( x_0 \) that is preserved at timestep \( t \). At each timestep $t\in(1,T)$, the noisy sample $x_t$ is computed as:

\begin{equation}
x_t = \sqrt{A_t}\, x_0 + \sqrt{1 - A_t}\, \epsilon, \quad \epsilon \sim \mathcal{N}(0, I)
\end{equation}
where $\epsilon$ is a noise sample drawn from a standard Gaussian distribution (analogous to $z$ in the VAE formulation above).

The reverse process aims to recover the clean spectorgram $x_0$ from the noise sample $x_T$ by progressively denoising over $T$ timesteps.  A neural network, typically a U-Net \citep{ronneberger2015u}, is trained to predict the added noise at each timestep. The U-Net architecture is particularly well-suited for this task due to its ability to capture hierarchical features at different resolutions. During training, the network predicts the noise $\epsilon$ at each timestep, minimizing the simplified ($\mathcal{L}_{\text{simple}}$) loss function:

\begin{equation}
\mathcal{L}_{\text{simple}} = \mathbb{E}_{t,x_0,\epsilon}\left[\left\|\epsilon - \epsilon_\theta\left(x_t,\, t\right)\right\|^2\right]
\end{equation}
where $\mathcal{L}_{\text{simple}}$ represents the mean-squared error (MSE) between the actual noise $\epsilon$ and the noise predicted by the network $\epsilon_\theta\left(x_t,\, t\right)$.  

As the model is trained to minimize the MSE loss function across many noisy samples, it gradually becomes better at removing the noise step by step, encouraging the network to become better at removing noise from the spectrogram and reconstructing the clean spectrogram, $x_0$. Once trained, the model synthesizes novel spectrograms by iteratively denoising random noise over $T$ timesteps.

DDPMs overcome many limitations of GANs and VAEs. Rather than relying on adversarial training or latent reconstruction, they minimize a denoising objective at each timestep, leading to more stable training and higher sample quality and diversity \citep{cao2024survey}. As a result, DDPMs are capable of generating highly detailed samples that often surpass those produced by both GANs and VAEs \citep{ho2020denoising, dhariwal2021diffusion}.

\subsection{Filtering Low-Quality Synthetic Spectrograms}

\begin{figure}[ht]
    \centering
    \includegraphics[width=0.25\linewidth]{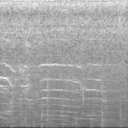}
    \includegraphics[width=0.25\linewidth]{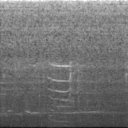}
    
    \vspace{0.3cm}
    
    \includegraphics[width=0.25\linewidth]{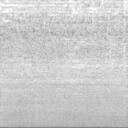}
    \includegraphics[width=0.25\linewidth]{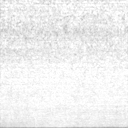}
    
    \caption{Examples of synthetic spectrograms generated using DDPMs. The top row shows samples that were accepted for training, exhibiting clear SRKW-like vocal structure. The bottom row shows samples that are unsuitable due to the presence of artifacts or poor signal definition.}
    \label{fig:gen_failures}
\end{figure}

Despite the success of DDPMs, and generative models in general, in producing visually appealing and realistic images, several limitations remain, particularly when such models are used for data augmentation in classification tasks. Generative models, including GANs and DDPMs, may synthesize samples containing artifacts or failure regions that can negatively impact downstream classifier performance. This issue can be especially pronounced in the context of marine bioacoustics, where spectrograms often contain low signal-to-noise ratios and complex background interference (e.g., vessel noise or other environmental sounds). In such cases, generative models may inadvertently learn to replicate the noise rather than the vocalization signal of interest. Figure~\ref{fig:gen_failures} showcases examples of common failures in synthetic spectrograms generated from noisy marine mammal datasets.

To mitigate the risk of low-quality or out-of-distribution synthetic samples negatively affecting the classifier model training, we implemented a simple PCA-based filtering strategy designed to align the statistical distribution of generated spectrograms with that of real SRKW data. 

Let $X_{real}$ and $X_{gen}$ denote the sets of real and generated spectrograms, respectively. To assess the quality of synthetic data, we applied PCA to $X_{real}$ to define a low-dimensional projection space, and projected both $X_{real}$ and $X_{gen}$ into this space. We then used the Mahalanobis distance to quantify how closely the projection of a generated spectrogram $z_{gen}$ aligned with the distribution of $X_{real}$ in this space.

A generated spectrogram was retained if its Mahalanobis distance $d_M$ satisfied:

\begin{equation} d_M(z_{\text{gen}}) \leq \tau \end{equation}
where $\tau$ is a threshold set to the $j$-th percentile of Mahalanobis distances computed from $X_{real}$. 

We applied this filtering procedure to the output of all generative models, to ensure that only synthetic samples statistically aligned with the distribution of real SRKW vocalizations were used for data augmentation.

\subsection{Deep Learning Classifier}
\label{sec:NN}

Our aim is to generate synthetic vocalizations using the augmentation methods described in the previous sections, and to use these data to train a DNN to detect SRKW calls in novel acoustic environments within a prescribed level of accuracy. To this end, we trained a convolutional neural network (CNN) to classify spectrograms as containing SRKW vocalizations, or not.

We employed a ResNet-18 architecture \citep{he2016deep}, a compact variant of the residual network family. Given the limited size of our training dataset the ResNet-18 strikes a balance between representational capacity and computational efficiency. The network is designed to accept 3-second Mel-spectrogram representations as input, and to output a binary classification probability indicating the presence or absence of SRKW vocalizations.

While the ResNet-18 was the architecture of choice in this study, numerous deep learning classifiers have demonstrated success in detecting and classifying marine mammal sounds. Classical CNN architectures, such as LeNet and VGG-16, as well as GRU RNNs, have shown robust performance in classifying North Atlantic Right Whale (NARW) vocalizations in early deep learning applications in marine bioacoustics \citep{shiu2020deep}. More recent architectures, such as DenseNets and Inception models have further advanced performance in related tasks \citep{10270176}. Indeed, the ResNet architecture itself has been widely adopted in the field \citep{padovese2021data, 9943551}, including for Killer Whale classification off the west coast of Canada \citep{bergler2019orca}. While our focus was not on identifying the optimal classifier, it is reasonable to expect that alternative architectures could also provide satisfactory performance based on these prior successes. Ultimately, the central challenge in bioacoustics remains addressing the constraints posed by complex noise environments and small annotated datasets rather than selecting the perfect neural network architecture.

\subsection{Experimental Setup}
\label{sec:expSetup} 

To evaluate the effectiveness of the data augmentation strategies, we conducted a series of experiments comparing classifier performance across seven training regimes: (I) baseline (no augmentation), (II) time-shifting augmentation only, (III) vocalization mask augmentation only, (IV) VAE-generated synthetic data only, (V) GAN-generated synthetic data only, (VI) DDPM-generated synthetic data only, and (VII) hybrid augmentation combining time-shifting, vocalization masks, and the best-performing generative model, selected empirically based on evaluation performance (see Section~\ref{sec:results}). Models were trained using vocalizations from the Lime Kiln dataset and evaluated on vocalizations from the independent Robert’s Bank dataset, allowing us to assess generalization to a site not used for training.

All experiments were carried out on a dedicated workstation equipped with an NVIDIA GeForce GTX 1070ti GPU ($8$ GB memory). In the following, we will describe the model architectures and parameters used for training the different models.

\paragraph{\textbf{Classifier}}

Each classifier was trained using the same ResNet-18 architecture described in Section~\ref{sec:NN}, trained for $20$ epochs with a batch size of $128$ samples. The model was optimized using an Adam optimizer with a learning rate of 0.001, and a cosine annealing scheduler with linear warmup. As the generative models operated on $128 \times 128$ spectrograms, all input data were resized to $128 \times 128$ when necessary, and normalized to $[0, 1]$.

\paragraph{\textbf{VAE}}

The VAE model consisted of a convolutional encoder with three layers (32, 64, 128 filters, kernel size 4, stride 2), each followed by ReLU activation, and two fully connected layers to compute the mean and log-variance of a 32-dimensional latent vector. The decoder mirrored this structure with a fully connected layer followed by three transposed convolutional layers (128, 64, 32 filters), each with ReLU activation, and a final output layer with tanh activation. Models were trained for 150 epochs using a batch size of 64, and a learning rate of 0.001 with an Adam optimizer. The loss combined MSE reconstruction and KL divergence. Input images were resized to $128 \times 128$ pixels and normalized to $[-1, 1]$.

\paragraph{\textbf{GANs}}

The GAN model followed a DCGAN-style architecture, with a generator composed of six transposed convolutional layers (with 1024, 512, 256, 128, 64, and 1 filters, respectively), each followed by BatchNorm and ReLU activations, except for the final layer which used tanh. The discriminator consisted of five convolutional layers with LeakyReLU activations and PhaseShuffle layers \citep{donahue2018adversarial} to promote invariance to local time shifts. Batch normalization was applied after all but the first convolutional layer. Both networks were trained using the Adam optimizer with a learning rate of 0.0001 and $(\beta_1, \beta_2) = (0.5, 0.999)$, for 300 epochs with a batch size of 64. Input images were resized to $128 \times 128$ pixels and normalized to the $[-1, 1]$ range.

\paragraph{\textbf{DDPM}}

For diffusion-based synthesis we adopted a standard DDPM built around a U‑Net backbone. 
The U‑Net processes $128 \times 128$ single‑channel spectrograms and comprises six resolution levels, with two residual blocks per level. The encoder used six blocks with 128, 128, 256, 256, 512, and 512 filters, respectively. The fifth block, operating at $8 \times 8$ resolution, incorporated self-attention to better capture long-range time–frequency structure, while the remaining blocks used standard downsampling operations. The decoder mirrored this structure in reverse.

We used the original DDPM formulation with 1,000 diffusion steps and a cosine $\beta_t$ schedule, training for 150 epochs with a batch size of 32. Optimization employed an AdamW optimizer (learning rate $1 \times 10^{-4}$, $\beta_1{=}0.9$, $\beta_2{=}0.999$) and an exponential moving average of the network weights (decay $0.9999$) for evaluation. During training the objective was the MSE between predicted and true noise. All inputs were resized to $128 \times 128$ pixels and normalized to the $[-1, 1]$ range, matching the other generative models for consistent downstream comparison.

\paragraph{\textbf{Experimental Protocol}}

The baseline model in experiment I was trained using only real data, that is, 1,261 3-second SRKW vocalization clips extracted from annotated regions and an equal number of background segments sampled from non-annotated intervals (Section~\ref{sec:data-prep}). This model establishes a performance reference point against which all augmented variants were compared.

For Experiments II through VII, augmented vocalization samples were generated independently and added to the baseline training set to create expanded datasets. In each case, we augmented the number of vocalizations by two different amounts, $+5{,}000$ and $+20{,}000$ samples, to investigate how classifier performance is affected by the quantity of synthetic data introduced during training. For Experiment VII, the hybrid strategy combined time-shifting, vocalization masking, and the best-performing generative model. To keep the total number of augmented samples consistent with the other experiments, the same augmentation budget of $+5{,}000$ and $+20{,}000$ new samples was used. This total was divided equally among the three augmentation methods, resulting in $+1{,}667$ new samples per method for the $+5{,}000$ setting and $6{,}667$ per method for the $+20{,}000$ setting. A summary of the experimental combinations and vocalization sample sizes is presented in Table \ref{tab:experiment_summary}.

To maintain class balance across all training sets, each set of vocalization samples, whether real or augmented, was paired with an equal number of background samples. In the baseline and time-shifting experiments (I and II), background segments were randomly drawn from non-annotated intervals in the Lime Kiln recordings. For the remaining experiments (III–VII), background samples were generated using the same strategy as their corresponding vocalizations. In the case of vocalization masking (Experiment III), since the masks were overlaid onto background spectrograms from the Robert’s Bank dataset (see Section~\ref{sec:masking}), we randomly sampled additional background segments from Robert’s Bank recordings that were outside the recordings reserved for the test set. 

For all synthetic augmentation methods (Experiments IV–VII), a separate generative model of the same type as the one used for vocalizations was trained exclusively on background data from Lime Kiln. For example, background samples in Experiment IV were generated using a VAE trained on background spectrograms, while Experiment V used a background GAN, and so forth. This approach ensured that both positive and negative samples in each experiment were drawn from comparable distributions and generated under consistent modeling assumptions. In light of this, throughout the text, when we refer to experiments being augmented by $5{,}000$ or $20{,}000$ vocalization samples, it should be understood that an equal number of background samples was also added, following the corresponding method for that experiment.
 
\begin{table*}[]
\centering
\small
\setlength{\tabcolsep}{4pt}
\caption{Summary of experiment combinations and number of vocalization samples. A dash indicates that no samples from that augmentation strategy were included. For Experiment VII, only the best-performing generative model was used. Each set of vocalizations was paired with an equal number of background samples generated using the same method.}
\label{tab:experiment_summary}
\begin{tabular}{lcccccc}
\toprule
Exp. & Real & Time-Shift & Masks & VAE & GAN & DDPM \\
\midrule
I (\textit{Baseline})   & 1,261 & –         & –         & –         & –         & –         \\
\midrule
II (\textit{Time-Shift})  & 1,261 & \makecell{+5,000\\+20,000}  & –         & –         & –         & –         \\
III (\textit{Masks}) & 1,261 & –         & \makecell{+5,000\\+20,000}  & –         & –         & –         \\
\midrule
IV (\textit{VAE}) & 1,261 & –         & –         & \makecell{+5,000\\+20,000}  & –         & –         \\
V  (\textit{GANS}) & 1,261 & –         & –         & –         & \makecell{+5,000\\+20,000}  & –         \\
VI (\textit{DDPM}) & 1,261 & –         & –         & –         & –         & \makecell{+5,000\\+20,000}  \\
\midrule
VII (\textit{Hybrid}) & 1,261 & \makecell{+1,667\\ +6,667} & \makecell{+1,667\\ +6,667} & – & – & \makecell{+1,667\\ +6,667} \\
\bottomrule
\end{tabular}
\end{table*}

Each experiment was repeated $10$ times with different random initializations. Performance was quantified using precision, recall, and F1-score, with final metrics averaged across all runs to establish the mean performance. The code used in this study has been made publicly available at \href{https://github.com/bpadovese/GAugSRKW}{https://github.com/bpadovese/GAugSRKW}, accompanied by comprehensive documentation and a command-line interface to facilitate reuse and adaptation for other bioacoustics applications.

\section{Results}
\label{sec:results}

\begin{figure*}[t]
    \centering

    \begin{tikzpicture}
        \node[anchor=north west, inner sep=0] at (0,0) {\includegraphics[width=0.60\linewidth]{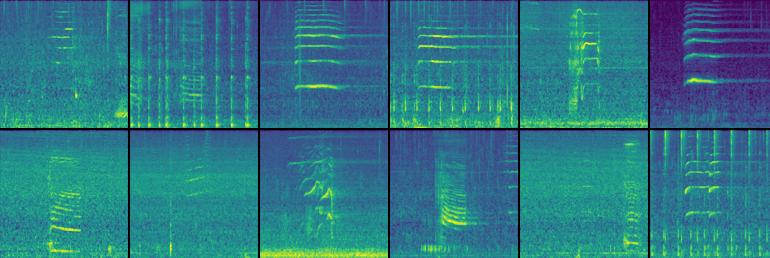}};
        \node[anchor=north west] at (0, -0.01) {\footnotesize\textbf{(a)}};
    \end{tikzpicture}

    \vspace{0.5em}
   
    \begin{tikzpicture}
        \node[anchor=north west, inner sep=0] at (0,0) {\includegraphics[width=0.60\linewidth]{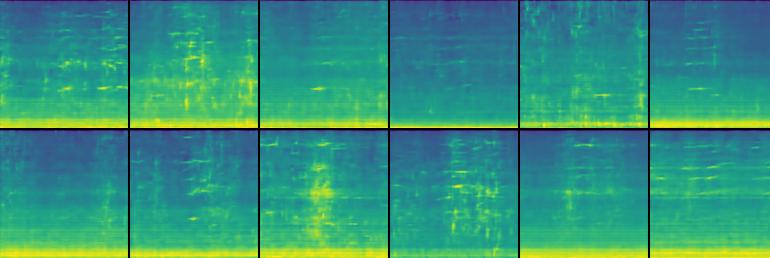}};
        \node[anchor=north west] at (0, -0.01) {\footnotesize\textbf{(b)}};
    \end{tikzpicture}

    \vspace{0.5em}
  
    \begin{tikzpicture}
        \node[anchor=north west, inner sep=0] at (0,0) {\includegraphics[width=0.60\linewidth]{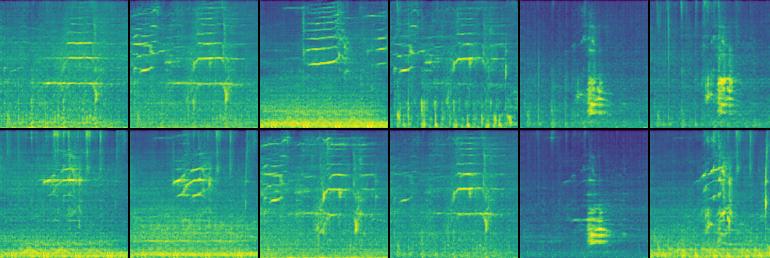}};
        \node[anchor=north west] at (0, -0.01) {\footnotesize\textbf{(c)}};
    \end{tikzpicture}
   
    \vspace{0.5em}
 
    \begin{tikzpicture}
        \node[anchor=north west, inner sep=0] at (0,0) {\includegraphics[width=0.60\linewidth]{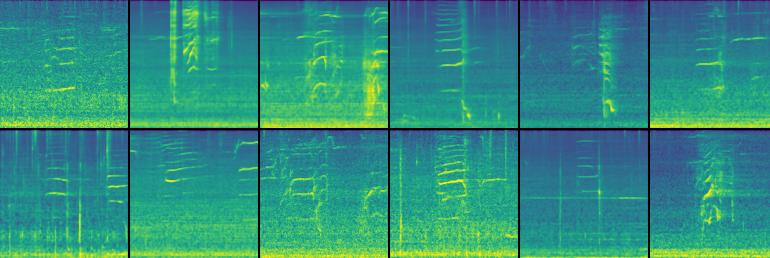}};
        \node[anchor=north west] at (0, -0.01) {\footnotesize\textbf{(d)}};
    \end{tikzpicture}
    
    \caption{Examples of (a) real SRKW vocalizations, (b) VAE-generated samples, \\(c) GAN-generated samples, and (d) DDPM-generated samples.}
    \label{fig:gen_comparison}
\end{figure*}

We illustrate a visual comparison between 12 real SRKW vocalization spectrograms in Figure~\ref{fig:gen_comparison} (a), and  the same number of synthetic samples generated by VAE (b), GAN (c), and DDPM (d) respectively. This subset is intended as a representative but small sample illustrating the qualitative differences among models. Visually, we observed that the VAE-generated samples suffered from a characteristic over-smoothing effect. Harmonic ridges appear blurred, and in many cases, signal boundaries are poorly defined or diffused into the background. 

The GAN samples, by contrast, displayed sharper structures and better-defined signals than the VAE outputs. However, despite avoiding the most severe cases of mode collapse, the GAN still seemed to produce a limited range of variation. Several spectrograms appear overly similar in composition, with vocalizations often occurring in the same time-frequency locations.

Finally, the DPPM synthetic spectrograms exhibited high visual fidelity, preserving fine harmonic structure and modulation contours that are characteristic of discrete pulsed calls, while avoiding the over-smoothed appearance often observed with VAEs and the low sample variety typical of GANs.

Table~\ref{tab:prf-aug-comparison} summarizes the classification performance in terms of precision, recall, and F1-score at a decision threshold of 0.5, a commonly adopted default in classification tasks used in many studies \citep{herbst2024empirical, paulphd, li2023learning}. Time-shifting augmentation (II) led to a substantial increase in recall, reaching 0.62 and 0.71 for the +5,000 and +20,000 settings, respectively, compared to just 0.42 for the baseline model. Precision remained comparable to the baseline at +5,000 samples (0.70 vs 0.65), but improved notably to 0.76 when the number of augmented samples increased to 20,000, resulting in F1-scores of 0.66 and 0.73, respectively. The mask-only strategy (III) consistently achieved near-perfect precision (0.98 and 0.99) at both augmentation levels, with overall comparable performance across settings. Recall remained relatively modest, rising slightly from 0.51 at +5,000 to 0.53 at +20,000.

The VAE-based augmentation (IV) did not provide substantial gains at the +5,000 augmentation level, with performance remaining comparable to the baseline (F1-score of 0.51). In contrast, the +20,000 setting resulted in a modest improvement, increasing the F1-score to 0.57. Similarly, GAN-based augmentation (V) produced comparable results, achieving F1-scores of 0.52 and 0.60. In contrast, the DDPM-based approach (VI) demonstrated consistently stronger performance, with F1-scores of 0.71 and 0.75, representing an improvement of approximately 0.15 over the other generative models at both augmentation levels. Notably, it achieved the highest recall overall, reaching 0.87 in the +20,000 setting. Finally, the combined augmentation approach (VII) delivered the best overall performance in the +20,000 setting, combining high precision at 0.99, with a recall of 0.69, resulting in the highest F1-score of 0.81.

\begin{table*}[]
\centering
\small
\caption{Precision, Recall, and F1-Score at Threshold 0.5 for each augmentation strategy. Experiment I serves as the baseline with 1,261 samples. Other experiments include augmented data with increasing sample sizes. Values are reported as mean $\pm$ standard deviation across ten runs. Bold values indicate the best performance within each augmentation level.}
\label{tab:prf-aug-comparison}
\begin{tabular}{llccc}
\toprule
\textbf{Experiment} & \makecell{\textbf{Number of}\\\textbf{Samples}} & \textbf{Precision} & \textbf{Recall} & \textbf{F1-Score} \\
\midrule
I (\textit{Baseline}) & 1,261  & 0.65 $\pm$ 0.024 & 0.42 $\pm$ 0.048 & 0.51 $\pm$ 0.034 \\
\midrule
II (\textit{Time-shift}) & + 5,000 & 0.70 $\pm$ 0.021 & 0.62 $\pm$ 0.030 & 0.66 $\pm$ 0.017 \\
   & + 20,000 & 0.76 $\pm$ 0.020 & 0.71 $\pm$ 0.021 & 0.73 $\pm$ 0.012 \\
\midrule
III (\textit{Masks}) & + 5,000 & 0.98 $\pm$ 0.004 & 0.51 $\pm$ 0.250 & 0.67 $\pm$ 0.021 \\
    & + 20,000 & \textbf{0.99} $\pm$ 0.003 & 0.53 $\pm$ 0.030 & 0.69 $\pm$ 0.025 \\
\midrule
IV (\textit{VAE}) & + 5,000 & 0.69 $\pm$ 0.033 & 0.41 $\pm$ 0.057 & 0.51 $\pm$ 0.039\\
   & + 20,000 & 0.78 $\pm$ 0.039 & 0.45 $\pm$ 0.024 & 0.57 $\pm$ 0.019 \\
\midrule
V (\textit{GAN}) & + 5,000 & 0.71 $\pm$ 0.025 & 0.41 $\pm$ 0.037 & 0.52 $\pm$ 0.026\\
  & + 20,000 & 0.76 $\pm$ 0.032 & 0.50 $\pm$ 0.051 & 0.60 $\pm$ 0.034\\
\midrule
VI (\textit{DDPM}) & + 5,000 & 0.70 $\pm$ 0.012 & 0.71 $\pm$ 0.040 & 0.71 $\pm$ 0.019\\
  & + 20,000 & 0.67 $\pm$ 0.029 & \textbf{0.87} $\pm$ 0.024 & 0.75 $\pm$ 0.010 \\
\midrule
VII (\textit{Hybrid})  & + 5,000 & 0.96 $\pm$ 0.011 & 0.63 $\pm$ 0.033 & 0.76 $\pm$ 0.021\\
& + 20,000 & \textbf{0.99} $\pm$ 0.003 & 0.69 $\pm$ 0.024  & \textbf{0.81} $\pm$ 0.015 \\
\bottomrule
\end{tabular}
\end{table*}

\begin{figure}[]
    \centering
    \includegraphics[width=\linewidth]{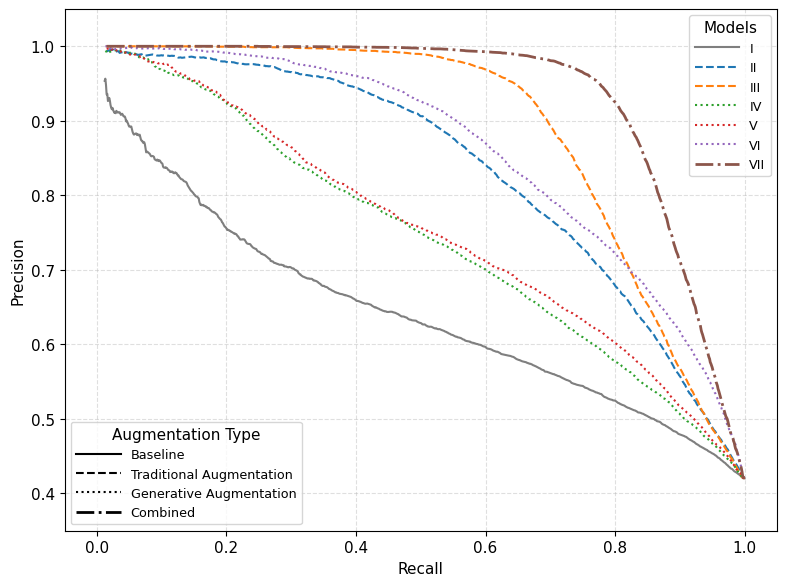}
    \caption{Mean Precision-Recall curves for all seven training regimes under the 20,000 augmentation setting, evaluated on the Robert’s Bank test set across the entire range of thresholds. Each curve shows the average performance over ten independent runs. The gray solid line represents the baseline model (I). Models II (blue dashed) and III (orange dashed) represent time-shifting and vocalization masking augmentations. Models IV (green dotted) and V (red dotted) and VI (purple dotted) use generative approaches based on VAE, GANs and DDPM, respectively. Finally, Model VII (brown dash-dot) combines the time-shift, masks and DDPM strategies.}
    \label{fig:pr_curves}
\end{figure}

Next, we analyzed the classification performance of the proposed methodology across all seven experimental regimes under different decision thresholds. Figure~\ref{fig:pr_curves} displays the mean precision-recall curves for all seven experiments described in Table~\ref{tab:experiment_summary}, showing classification performance with respect to the SRKW class for the +20,000 augmentation setting. Each curve represents the average over ten training runs. The gray line corresponds to the baseline model (I), trained exclusively on real vocalizations. The blue and orange lines represent time-shifting (II) and vocalization mask (III) augmentations, respectively. The green, red, and purple lines correspond to generative augmentation strategies using VAE (IV), GAN (V), and DDPM (VI). Finally, the brown line (VII) shows the performance of the combined approach using time-shifting, masking, and DDPM-generated samples.

Unlike Table \ref{tab:prf-aug-comparison}, which reports precision, recall, and F1-score at a fixed decision threshold of 0.5, the precision–recall curves in Figure \ref{fig:pr_curves} illustrate performance across the entire range of thresholds, and provide a more complete picture of model behavior. The baseline model (I) achieved the lowest performance across the board, with steep precision decay at low recall levels, reflecting the limited generalization when trained only on a small set of real samples from Lime Kiln. As expected, applying any augmentation method improved performance to varying degrees. Traditional augmentation methods maintained precision above 0.9 for recall values below 0.5, with the vocalization mask model (III) outperforming time-shifting (II) across most of the curve. Notably, the mask-only model achieved near-perfect precision at the cost of recall, showing a sharp decline as recall increased above 0.6.

Among the generative approaches, VAE-based (IV) and GAN-based (V) augmentation produced similar precision–recall profiles, both outperforming the baseline but lagging behind traditional augmentation strategies by more than $\sim$0.1 in precision across most of the curve. In contrast, the DDPM model (VI) consistently outperformed not only the VAE and GAN models but also the time-shifting approach (II) across the entire recall range. We note, however, that the DDPM model only surpassed the vocalization masking model (III) at the upper end of the recall spectrum (above $\sim$0.8), where it maintained higher precision.

Finally, the best overall performance was achieved by the combined strategy (VII), which consistently maintained the highest precision for any recall. The improvement was especially pronounced at higher recall values, with near-perfect precision remaining stable up to recall $\sim$0.7 and only dropping below 0.9 beyond $\sim$0.8. Even at these higher recall levels, the model preserved a margin of $\sim$0.2 in precision compared to all other strategies, highlighting the improved generalization to the independent test site.

\section{Discussion}
\label{sec:disc}

The performance of deep learning models in bioacoustic classification tasks is strongly dependent on the availability and diversity of annotated training data. To address this limitation without relying on additional expensive and time consuming manual labeling, data augmentation is often employed to expand the effective size and variability of training sets. In this work, traditional data augmentation techniques were first evaluated for their ability to enhance model performance. The baseline model, trained exclusively on real SRKW vocalizations from the Lime Kiln dataset, performed poorly when tested on the independent Robert’s Bank site. This outcome was expected given the acoustic variability between locations and differences in hydrophones. Differences in background noise profiles and sound propagation conditions often limit a model’s ability to generalize beyond its training domain. In contrast, augmenting the training data with time-shifted (experiment II) versions of the original vocalizations led to a substantial performance boost over the baseline model, highlighting its utility in scenarios where annotated vocalization samples are scarce. Time-shifting is computationally inexpensive, easy to implement, and resulted in consistently higher precision–recall performance compared to the baseline. Other lightweight augmentation methods such as pitch shifting \citep{white2022more, paulphd}, additive noise \citep{white2022more}, and random masking \citep{park2019specaugment}, have shown similarly measurable improvements across various bioacoustic applications.

More sophisticated approaches, such as vocalization mask augmentation (experiment III), further improved model performance by enabling context-aware signal insertion. By overlaying curated, high-quality vocalization templates onto real background noise from the target environment this method generated realistic and acoustically plausible training samples. Conceptually, this approach relates to prior work by \cite{li2020learning}, who synthesized dolphin whistle contours for model training by blending artificial whistle shapes with natural background spectrograms. However, our implementation differs by using empirically recorded SRKW vocalizations as masks, rather than synthetic contours, yielding ecologically grounded composites that preserve the spectral characteristics of genuine calls. Crucially, it exposed the model to the specific environmental context in which detection is expected to occur, allowing it to learn discriminative features that are directly relevant for deployment and can be observed by the drastic increase in precision, demonstrating the model’s improved ability to avoid false positives. However, because the curated call catalogue contained a limited number of vocalization types, and only high quality vocalizations, the model was not exposed to the full natural variability of SRKW vocalizations. In real-world conditions, such as those present in the Lime Kiln and Robert’s Bank datasets, vocalizations can often appear faint, overlapped, or embedded within complex acoustic environments. This is reflected by the more modest recall performance observed, indicating that the model failed to detect many calls due to its narrow exposure to vocalization diversity. Furthermore, increasing the number of augmented samples from +5,000 to +20,000 provided limited performance gains, suggesting that adding more homogeneous high-quality calls did not increase data variability or improve generalization.

Nonetheless, traditional augmentation techniques are ultimately limited by the scope and structure of the original dataset. There is only so much variability that can be meaningfully introduced through manual transformations. In this regard, generative approaches offer an alternative. Generative methods used in this study hold the potential to unlock new regions of the data distribution by synthesizing entirely novel yet realistic samples. All generative methods outperformed the baseline, indicating their ability to synthesize relevant and in-distribution samples that enhanced classifier performance. While both the VAE-based and GAN-based approaches underperformed traditional augmentation methods, they exhibited a more gradual decline in precision as recall increased. This pattern suggests that, despite some imperfections, generative models captured a broader spectrum of the vocalization space, including samples that were more challenging or ambiguous. 

This trend is most evident with the DDPM-based approach, which outperformed the other generative models and time-shifting across the entire recall range, and surpassed all augmentation strategies at the upper end of the recall spectrum. More importantly, combining DDPM-generated samples with traditional augmentations led to the highest overall classification performance, exceeding what any single strategy could achieve in isolation. This hybrid configuration appears to be a promising strategy for bioacoustic applications, as it leverages the complementary strengths of generative and non-generative augmentations. The improvement is particularly notable when compared to the mask-only augmentation experiment where both the hybrid and mask-only models maintained similarly high precision at low recall levels, but only the hybrid approach sustained that precision as recall increased. These results suggest that generative and non-generative augmentations can complement each other. The former expands the diversity of vocalization forms beyond the limits of the original dataset, while the latter ensures contextual realism. Their combination enables the model to generalize more effectively by learning from a richer set of training scenarios.

Despite the promising results, several limitations must be acknowledged. Most notably, the use of generative models such as GANs, VAE, and DDPM in particular, introduces significant computational overhead. Training these models requires substantial GPU resources and time, particularly when compared to traditional augmentation techniques like time-shifting or template overlay, which are fast, lightweight, and easy to implement. Inference with DDPMs can also be relatively slow, especially for generating high quality samples, further limiting their practicality in real-time or large-scale training workflows. Additionally, contrary to CNN-based classifiers, which are now well established and supported by numerous off-the-shelf implementations, generative methods demand a certain degree of domain knowledge to implement and apply effectively. This higher complexity could limit their adoption, particularly among practitioners without specialized expertise. Moreover, while DDPM-based augmentation improved overall performance when used in combination with other methods, it did not consistently outperform conventional techniques when used in isolation. These findings broadly correspond with the work of \cite{herbst2024empirical}, who reported that VAE and DDPM-based augmentation did not substantially improve classifier performance of Hainan gibbon calls beyond what could be achieved with a suite of traditional augmentation methods. This suggests that the added complexity may not always be justified in scenarios where computational resources are constrained or rapid deployment is required. Thus, there is a clear tradeoff between the potential gains in data diversity offered by generative models and the practical demands of training and integrating them into existing pipelines.

Nonetheless, the ability of generative models to produce a potentially unlimited number of diverse samples, or at least far beyond what simpler augmentation methods can offer, holds significant promise for bioacoustic applications. The capacity to synthesize novel vocalization patterns, drawn from a real distribution, that still conform to the underlying structure of real vocalizations can complement training pipelines for models that need to generalize across ecotypes, acoustic environments, or rare call types.

\section{Conclusion}
\label{sec:conc}

In this work, we explored several data augmentation strategies to address the limitations of small, site-specific bioacoustic datasets and to improve generalization to new recording environments. Our results demonstrate that while traditional augmentation techniques such as time-shifting and vocalization masking offer tangible performance improvements with minimal computational overhead, generative models can further enrich training datasets and enhance model robustness. In contexts where expert annotation is difficult to obtain, or where vocalizations are rare due to limited accessibility (e.g., far offshore) or the scarcity of the species itself, these computational methods represent a justifiable investment, maximizing the utility of available data and improving overall model performance.

This study serves as a stepping stone toward the broader integration of generative models into bioacoustic data augmentation pipelines. While these models introduce added complexity and computational demands, our findings suggest that their inclusion in targeted augmentation strategies can be a worthwhile investment in scenarios where generalization and recall performance are critical. In particular, the combination of generative and traditional augmentation techniques appears to be a promising strategy for enhancing model robustness and cross-site generalization.

Future work could explore diffusion models that operate in a compressed latent space, where spectrograms are first encoded into a lower-dimensional representation before the generative process. This approach can dramatically reduce training and inference costs while preserving sample quality. Additionally, further research into quality control strategies, such as integrating filtering approach similarly to the a PCA-based strategy used in this study directly into the DDPM training process, could improve the reliability of generated samples and enhance the scalability and effectiveness of generative augmentation approaches.

\section*{Funding sources}

This research was supported by the Humans and Algorithms Listening and Looking for Orcas project (HALLO) funded by the Canada Nature Fund for Aquatic Species at Risk of Fisheries and Oceans Canada (2022-NF-PAC-593022).

\section*{Acknowledgments}

We would like to thank April Houweling, Lauren Laturnus and Dr Jennifer Wladichuck for their help in annotating Killer Whale vocalizations.

% \appendix

% \section{Sample Appendix Section}
% \label{sec:sample:appendix}
% Lorem ipsum dolor sit amet, consectetur adipiscing elit, sed do eiusmod tempor section \ref{sec:sample1} incididunt ut labore et dolore magna aliqua. Ut enim ad minim veniam, quis nostrud exercitation ullamco laboris nisi ut aliquip ex ea commodo consequat. Duis aute irure dolor in reprehenderit in voluptate velit esse cillum dolore eu fugiat nulla pariatur. Excepteur sint occaecat cupidatat non proident, sunt in culpa qui officia deserunt mollit anim id est laborum.

\bibliographystyle{elsarticle-harv} 
\bibliography{references}





% \end{thebibliography}
\end{document}